# OPTIMIZATION OF FUZZY ANALOGY IN SOFTWARE COST ESTIMATION USING LINGUISTIC VARIABLES

S.Malathi[a], Dr.S.Sridhar[b] a*

[a]Research Scholar, Dept of CSE, Sathyabama University, Chennai, India.

[b]Research Supervisor, Dept of CSE & IT, Sathyabama University, Chennai, India.

**Abstract**

One of the most important objectives of software engineering community has been the increase of useful models that beneficially explain the development of life cycle and precisely calculate the effort of software cost estimation. In analogy concept, there is deficiency in handling the datasets containing categorical variables though there are innumerable methods to estimate the cost. Due to the nature of software engineering domain, generally project attributes are often measured in terms of linguistic values such as very low, low, high and very high. The imprecise nature of such value represents the uncertainty and vagueness in their elucidation. However, there is no efficient method that can directly deal with the categorical variables and tolerate such imprecision and uncertainty without taking the classical intervals and numeric value approaches. In this paper, a new approach for optimization based on fuzzy logic, linguistic quantifiers and analogy based reasoning is proposed to improve the performance of the effort in software project when they are described in either numerical or categorical data. The performance of this proposed method exemplifies a pragmatic validation based on the historical NASA dataset. The results were analyzed using the prediction criterion and indicates that the proposed method can produce more explainable results than other machine learning methods.

## 1. Introduction

Software cost estimation has been the subject of intensive investigation in the field of software engineering. Generally, effort estimation for software projects fall into two main categories namely algorithmic and non algorithmic models [1]. Algorithmic estimation involves the application of mathematical computation method. Non algorithmic estimation is purely based on machine learning techniques. Software cost estimation by analogy is one of the most striking machine learning techniques and is basically a form of Case-Based Reasoning [2]. It is based on the following assumption: similar software projects have similar costs. There are two main advantages of analogy-based estimation: first, its process is easy to understand and explain to users; and, second, it can model a complex set of relationships between the dependent variables (such as cost or effort) and the independent variables (cost drivers). However, its deployment in software cost estimation still warrants some improvements in handling the categorical variables.

* S.Malathi, Tel: 09382160092
E-mail address: malathi_raghu@hotmail.com

Fuzzy logic based [3] cost estimation models are more suitable when vague and imprecision information is to be accounted for. The advantage of this method is that they are more natural and they are similar to the way in which the human interprets the linguistic values. Though many membership functions are used in literature, Gaussian function outperforms other membership functions. The drawback of fuzzy method is that the imprecision and uncertainty are not accounted resourcefully. Even for implementing the COCOMO technique [4], fuzzy logic method is used.

Wei Lin Du et al. [5] proposed an approach combining the neuro-fuzzy technique and the SEER-SEM effort estimation algorithm. Moreover, continuous rating values and linguistic values can be inputs of the proposed model for avoiding the large estimation deviation among similar projects. However, this method has not shown the direction of handling the dataset effectively to overcome the fuzzy logic problem.

The proposed method effectively estimates the software effort using analogy technique with the classical fuzzy approach based on reasoning by analogy and fuzzy logic to estimate effort when software projects are described by linguistic values, which is a major limitation of all estimation techniques such as 'very low', 'low' and 'high'.

The paper is divided into 5 sections as follows. Section 2 gives the problem definition with the related work. The key features of the Fuzzy Analogy approach are presented in section 3. This includes the basic concepts of Analogy, Fuzzy Logic and Fuzzy Analogy. In section 4, an explorative analysis is conducted for validating the proposed method and as a result, a refined Fuzzy Analogy approach with the outcomes is presented in section 5. The conclusion of the findings is dealt in Section 6.

## 2. Related Work

Several researchers have carried out researches in the field of effort estimation for the software projects using various techniques [6]. These techniques, dealing with a few of the significant researches, have been highlighted here for iris recognition. Estimation of effort can be carried out in an efficient and accurate manner by collecting relavent software data terms. For the collection of such data, agile methodology [7] can be employed which is an accurate, incremental and an iterative one. Mohamed Azzeh *et al*., [8] have improved the performance of analogy at the early stage of identification process by using fuzzy numbers. The early techniques for effort estimation were typically based on statistics and regression analysis. Theoretical maximum prediction accuracy (*TMPA*) is robust software metric for software cost estimation using analogy in addition to existing model performance criteria such as *MMRE*.

Another important aspect of effort estimation is the fuzzy logic approach. Ahmeda and Muzaffar [9] dealt with the imprecision and uncertainty in the inputs of effort prediction. The research presents a transparent, enhanced fuzzy logic based framework for software development effort prediction. The Gaussian MFs [10] used in the fuzzy framework have shown good results by handling the imprecision in inputs quite well and also their ability to adapt further make them a valid choice to represent fuzzy sets. The framework is adaptable to the changing environments and handles the inherent imprecision and uncertainties present in the inputs quite well. M.Kazemifard et al., [11] uses a multi agent system for handling the characteristics of the team members in fuzzy system. There are many studies that utilized the fuzzy systems to deal with the ambiguous [12] and linguistic inputs of software cost estimation. In [13], homogeneous dataset results in better and more accurate effort estimates while the irrelevant and chaotic dataset results in lesser accuracy in effort estimations.

The Evolutionary Parallel Gradient Search (EPGS) uses EA to keep the best tracks of multiple searches and updates the best ones with gradient method. The parallel gradient search starts multiple searching at different points simultaneously to increase the opportunity of finding the global minimum [14]. ECM [15] introduces fuzzy set to describe the attributes of events and activities. Experiments show that it allows for more tolerance of project uncertainties and improves estimation precisions.

As such, an improved Fuzzy Analogy technique is proposed that attempts to reduce the effort performance by utilizing the linguistic values, size, actual effort drivers basically depending on SLOC and FP to overcome the drawback of fuzzy method in regard with the imprecision and uncertainty as well as handling the categorical variables efficiently.

## 3. Proposed Work

This paper is oriented towards the background reasoning of analogy using fuzzy logic which is carried out in two steps. The initial step is the study of analogy concept for each dataset and the next step is to use the fuzzy concept for implementing the fuzzy analogy to the individual dataset for effective cost estimation.

### 3.1. Analogy

Analogy based effort estimation method belongs to machine learning category. The basic idea of analogy prediction [16] is shown in Fig.1.

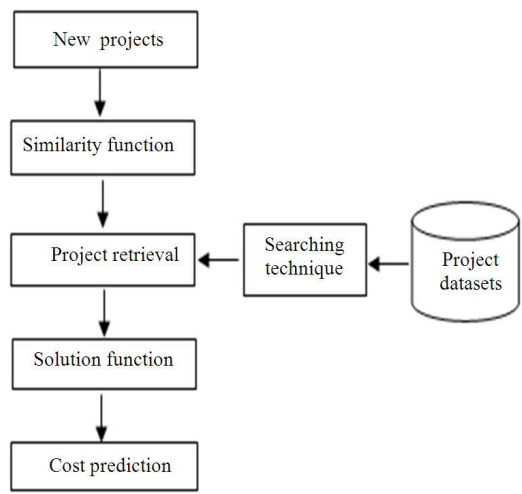

Fig.1 Concept of Analogy cost estimation

Projects that are similar with respect to project and product features such as size and complexity will be similar with respect to project effort".The steps involved are:

- Select the historic projects and find the cost drivers.
- Find the similarities between the new and the target projects.
- Recognize the historic projects that are analogous to the target.
- Set the effort of the historic project to estimate the effort of the target project.

The strong point of this method is that the estimate is based on actual project experience. However, it is not clear to what extent the previous project is actually representative of the limitation, situation and role to be performed by the new system.

To overcome this drawback, a new research called Analogy-X [17] that uses Mantel's correlation and randomization tests to verify the basic hypothesis of finding the numerical basis for analogy.This method removes the abnormal datapoints using the sensitivity analysis but still it does not handle the categorical datasets powerfully.

*3.2. Fuzzy logic*

Fuzzy logic is based on the human behaviour and reasoning. It has an affinity with fuzzy set theory and applied in situations where decision making is difficult. A Fuzzy set can be defined as an extension of classical set theory by assigning a value for an individual in the universe between the two boundaries that is represented by a membership function.

$$A = \int_x \mu_A(x)/x \tag{1}$$

Where x is an element in X and $\mu_A(x)$ is a membership function. A Fuzzy set is characterized by a membership function that has grades between the interval [0, 1] called grade membership function. There are different types of membership function, namely, triangular, trapezoidal, Gaussian etc.
Fuzzy logic consists of the following three stages:
        1. Fuzzification
        2. Inference Engine
        3. Defuzzification

The Fuzzifier transforms the inputs into a membership value for the linguistic terms. The function of inference engine is to develop the complexity matrix for producing a new linguistic term to determine the productivity rate by using fuzzy rules. A defuzzifier carries out the Defuzzification process to combine the output into a single label or numerical value as required.

*3.3. Fuzzy Analogy*

Fuzzification of classical analogy procedure is Fuzzy analogy. It comprises the following procedures, viz., 1) Identification of cases, 2) Retrieval of similar cases and 3) Case adaptation. Each step is the fuzzification of its equivalent classical analogy procedure.

*3.3.1 Identification of cases*

The goal of this step is the characterization of all software projects by a set of attributes. Selecting attributes, which describe software projects, is a complex task in the analogy procedure. Indeed, the selection of attributes depends on the objective of the CBR system. In this case, the objective is to estimate the software project effort. Consequently, the attributes must be relevant for the effort estimation task. The objective of the proposed Fuzzy Analogy approach is to overcome one of the drawbacks of analogy in handling the categorical variables and fuzzy method. Therefore, in the identification step, each software project is described by a set of selected attributes which can be measured by numerical or categorical values. These values will be represented by fuzzy sets. The framework is shown in figure 2.

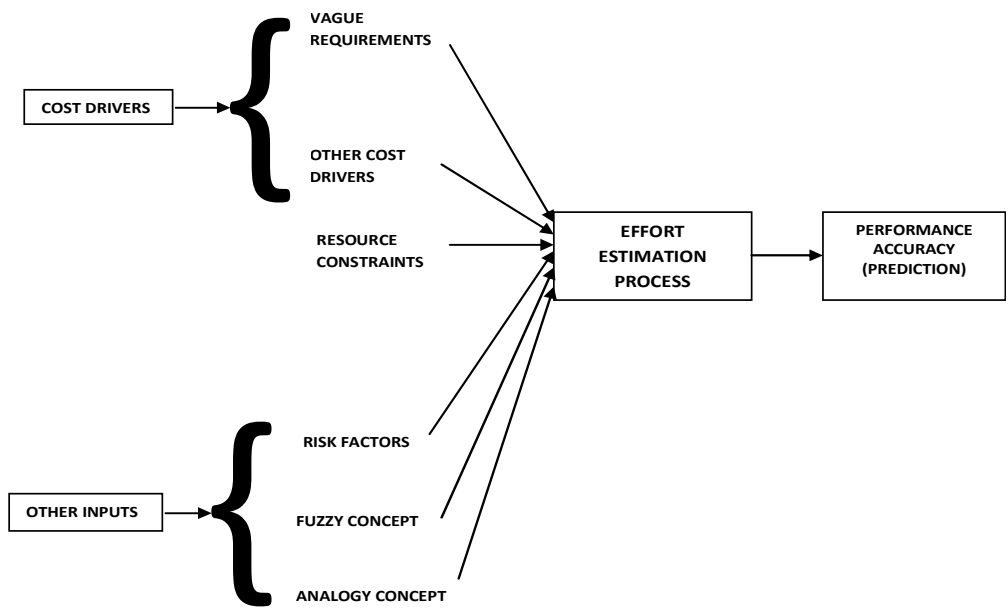

Fig.2 Framework of Fuzzy Analogy

In the case of numerical value $x_0$, its fuzzification will be done by the membership function which takes the value of 1 when $x$ is equal to $x_0$ and 0 otherwise. For categorical values, $M$ attributes are considered and for each attribute $M_j$, a measure with linguistic values is defined ($A_k^j$). Each linguistic value $A_k^j$ is represented by a fuzzy set with a membership function ($\mu_{A_k^j}$).

It is preferable that these fuzzy sets satisfy the normal condition. The use of fuzzy sets to represent categorical data, such as 'very low' and 'low', is similar to how humans interpret these values and consequently it allows dealing with imprecision and uncertainty in the case identification step.

*3.3.2 Retrieval of cases*

This step is based on the choice of software project similarity measure. In this method, a set of candidate measures for software project similarity has been proposed for software project similarity. These measures assess the overall similarity of two projects $P_1$ and $P_2$, $d(P_1, P_2)$ by combining all the individual similarities of $P_1$ and $P_2$ associated with the various linguistic variables $V_j$ describing the project $P_1$ and $P_2$, $d_{V_j}(P_1, P_2)$. After an axiomatic validation of some proposed candidate measures for the individual distances $d_{V_j}(P_1, P_2)$, two measures have been retained [18].

$$d_{V_j}(P_1, P_2) = \begin{cases} \max_k \min(\mu_{A_k^j}(P_1), \mu_{A_k^j}(P_2)) \\ max-min \ aggregation \\ \sum_k \mu_{A_k^j}(P_1) \times \mu_{A_k^j}(P_2) \\ sum-product \ aggregation \end{cases} \quad (2)$$

Where $A_k^j$ is the fuzzy set associated with $V_j$ and $\mu_{A_k^j}$ which are the membership functions representing fuzzy sets $A_k^j$. Scale factors (SF) are understanding product objectives, flexibility, team coherence, etc., Effort multipliers (EF) are software reliability, database size, reusability, complexity etc. The imprecision of the cost drivers significantly affects the accuracy of the effort estimates which are derived from effort estimation models. Since the imprecision of software effort drivers cannot be overlooked, a fuzzy model gains advantage in verifying the cost drivers by adopting fuzzy sets.

$$Effort = A * (SIZE)^{B + 0.01 * \sum_{i=1}^{N} d_i} * \prod_{i=1}^{N} EM_i \quad (3)$$

Where *A* and *B* are constants, d is the distance and EM is effort multipliers. By using the above formula, the effort is estimated. The cost drivers are fuzzified using triangular and trapezoidal fuzzy sets for each linguistic value such as very low, low, nominal, high etc. as applicable to each cost driver. Rules are developed with cost driver in the antecedent part and corresponding effort multiplier in the consequent part. The defuzzified value for each of the effort multiplier is obtained from individual Fuzzy Inference Systems after matching, inference aggregation and subsequent Defuzzification. Total Effort is obtained after multiplying them together. The high values for the cost drivers lead an effort estimate that is more than three times the initial estimate, whereas low values reduce the estimate to about one third of the original.

*3.3.3 Case adaptation*

The objective of this step is to derive an estimate for the new project by using the known effort values of similar projects. There are two issues that have to be addressed, (i) the choice of how many similar projects should be used in the adaptation, and (ii) how to adapt the chosen analogies in order to generate an estimate for the new project. In the available literature, it can be clearly noticed that there is no definite rule to guide the choice of the number of analogies. Fixing the number of analogies for the case adaptation step is considered here neither as a requirement nor as a constraint.

**4. Experimental Results**

This section explains the accuracy of effort estimation by the proposed work as well as the performance against other methods. The standard datasets are chosen from the available software engineering public domain as follows. In this method, NASA 93 [19] was selected consisting of 93 projects in various programming languages. The dataset are in COCOMO81 format collected from different NASA centres published in PRedictOR models in software engineering (PROMISE). The proposed work is implemented by using the default packages of JAVA Netbeans.

Table 1-3 and Figure 3-5 depict the actual and estimated effort against project ID based on simple, average and complex test sets. For each set, the estimation computed by a few selected cases is compared with the actual value of that case. For eg., plot of NASA93 dataset referring to simple test set is depicted in Figure 2, which shows that the estimated effort is efficient compared to the actual effort. Such idiosyncrasies can be regconized from 3 reasons, (1) elimination of abnormal data points; (2) Using the linguitics values based on analogy concept; (3) statiscally considering all the features even incase of additional features. The last reason is the major caveats that hinder the proposed work.

Table 1 Comparison of estimated and actual effort for simple projects in NASA93 dataset

| Project ID | Actual Effort | Estimated Effort |
|---|---|---|
| 1 | 117.6 | 115.89 |
| 3 | 31.2 | 29.764 |
| 4 | 36 | 34.144 |
| 5 | 25.2 | 24.232 |
| 6 | 8.4 | 8.829 |
| 7 | 10.8 | 11.028 |
| 9 | 72 | 66.623 |
| 10 | 72 | 66.10 |
| 11 | 24 | 23.204 |
| 15 | 48 | 42.621 |
| 18 | 60 | 52.594 |
| 20 | 60 | 52.830 |
| 40 | 114 | 97.723 |
| 48 | 252 | 214.87 |
| 49 | 107 | 91.635 |

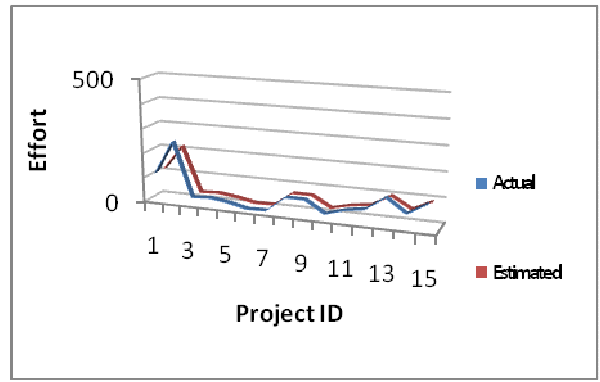

Fig.3 Comparison of Effort for simple projects in NASA93 dataset

Table 2 Comparison of estimated and actual effort for average projects in NASA93 dataset

| Project ID | Actual Effort | Estimated Effort |
|---|---|---|
| 8 | 352.8 | 324.538 |
| 12 | 360 | 322.68 |
| 14 | 215 | 191.545 |
| 16 | 360 | 305.99 |
| 17 | 324 | 277.815 |
| 22 | 300 | 256.11 |
| 38 | 444 | 378.77 |
| 41 | 1248 | 1060.761 |
| 45 | 400 | 340.088 |
| 50 | 571.4 | 484.976 |
| 53 | 750 | 636.722 |
| 54 | 2120 | 1797.690 |
| 56 | 1181 | 1001.664 |
| 57 | 278 | 235.629 |
| 63 | 162 | 138.343 |

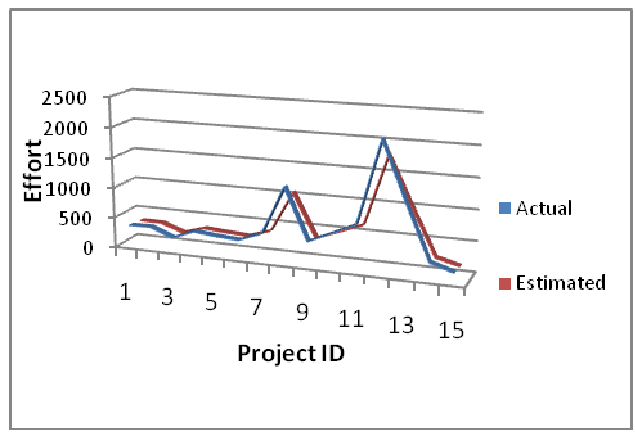

Fig.4 Comparison of Effort for average projects in NASA93 dataset

Table 3 Comparison of estimated and actual effort for complex projects in NASA93 dataset

| Project ID | Actual Effort | Estimated Effort |
|---|---|---|
| 42 | 2400 | 2035.135 |
| 43 | 1368 | 1160.729 |
| 44 | 973 | 825.677 |
| 46 | 2400 | 2035.818 |
| 59 | 4560 | 3864.849 |
| 60 | 720 | 610.667 |
| 62 | 2460 | 2082.503 |
| 67 | 444 | 376.429 |
| 77 | 1200 | 1015.503 |

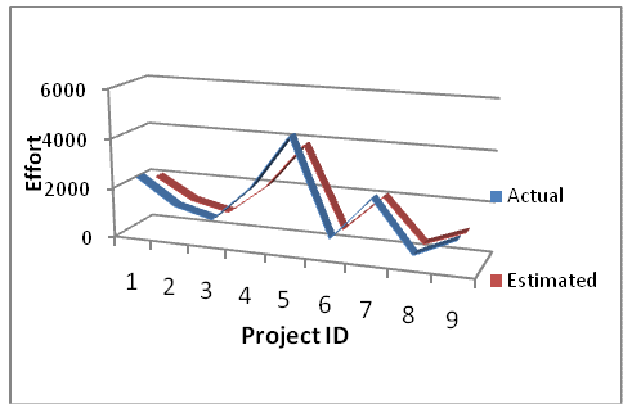

Fig.5 Comparison of Effort for complex projects in NASA93 dataset

The overall analysis of the projectset based on complexity performance is given in figure 6.

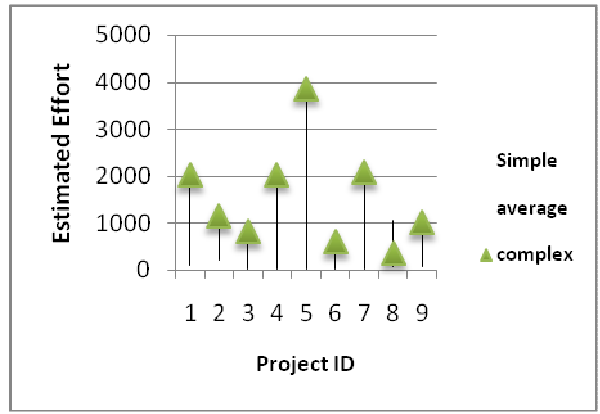

Fig.6 Comparison of overall complexity analysis in NASA93 dataset

## 5. Comparative Results

To assess the accuracy of the Fuzzy Analogy method, the common evaluation criteria is used in the field of software cost estimation. Prediction PRED (*p*) which represents the percentage of MRE that is less than or equal to the value p among all projects. This measure is often used in the literature and is the proportion of the projects for a given level accuracy [20]. The definition of PRED (*p*) is given as follows:

$$PRED(p) = \frac{k}{N}$$

Where *N* is the total number of observations and k is the number of observations whose MRE is less or equal to p. A common value for *p* is 25, which is used in the present study. The relative error of the proposed method is compared with the existing methods of COCOMO and Fuzzy model is shown in Figure 7.

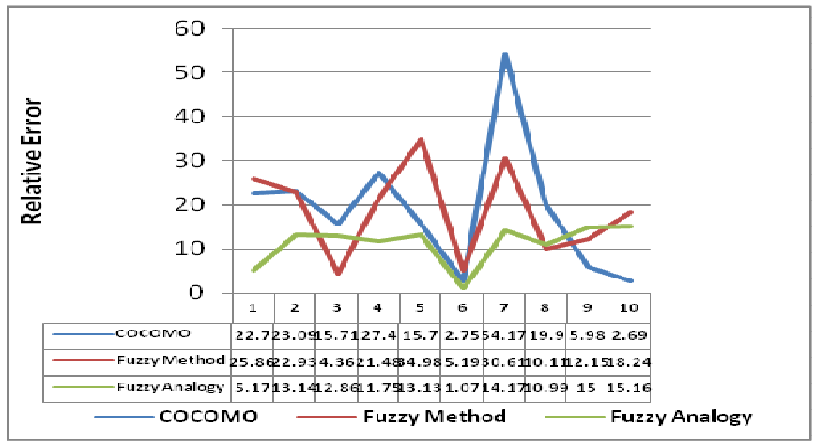

Fig.7 Comparison of Relative Error for NASA93 dataset with the existing methods

The overall comparative results of the PRED are tabulated for the NASA93 dataset in Table 4.

Table 4 Comparative Results of Prediction Accuracy

| Dataset | PRED(0.25) |
|---|---|
| Proposed Method | 0.86 |
| Fuzzy Method | 0.81 |
| COCOMO Method | 0.52 |
| CBR Method | 0.35 |

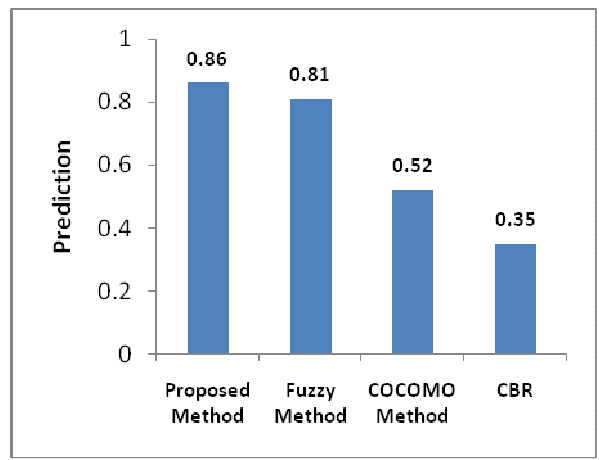

Fig.8 Comparison of Prediction for NASA93 dataset

From Figure 8, it is clearly observed that the proposed fuzzy analogy outperforms the accuracy achieved with Fuzzy, COCOMO [21] and CBR [8] methods.

## 6. Conclusion

Fuzzy Analogy is the most frequently applied method for cost estimation. This paper presents a holistic approach to achieve better results while handling the linguistic variables. Fuzzy analogy is compared against COCOMO, CBR and Fuzzy method and it is adduced that Fuzzy Analogy can achieve better predictions than the other soft computing methods. The results denote that the proposed method could be a promising improvement in the context of cost estimation. However, the results presented in this paper are preliminary because this method is validated with only one dataset and compared with the existing methods. In order to achieve optimum results, fuzzy analogy should be compared with datasets containing both the numerical and categorical variables. It is imperative and invariably essential to apply this method with the real projects to further understand the system of operation under varied conditions.